\input harvmac.tex
\vskip 2in
\Title{\vbox{\baselineskip12pt
\hbox to \hsize{\hfill}
\hbox to \hsize{\hfill IHES/97/P/23}}}
{\vbox{\centerline{S-Duality as an Open String Gauge Symmetry}}}
\centerline{Dimitri Polyakov\footnote{$^\dagger$}
{polyakov@ihes.fr}}
\medskip
\centerline{\it Institut des Hautes Etudes Scientifiques}
\centerline{\it 35, Route de Chartres, F-91440 Bures-sur-Yvette, FRANCE}
\vskip .5in
\centerline {\bf Abstract}
We demonstrate and
discuss the connection between S-duality  and string-theoretic
picture-changing gauge transformation in the ``infinite momentum frame''.
The picture-changing transformation at zero momentum
leads to the presence of the topological 5-form charge
in the type IIB superalgebra (dimensionally reduced
SUSY algebra of M-theory)
which is attributed to
the M5-brane.
The topological charge defines the 
D5-brane  state, which also is the analogue of the
monopole part of the Olive-Witten's 
result in field theory.The correlation functions involving
this monopole-like state are computed.
The fivebrane states are 
associated with the ghost number cohomologies which we introduce 
in the paper.
\Date{February 97}
\vfill\eject
\lref\sez{E.Sezgin, hep-th/9609086, CTP-TAMU-27-96}
\lref\sezg{E.Sezgin, hep-th/9612220, CTP-TAMU-78-96}
\lref\shenker{D.Friedan,S.Shenker,E.Martinec,Nucl.Phys.{\bf B271}(1986) 93}
\lref\sezgi{P.S.Howe, E.Sezgin, P.C.West, KCL-TH-97-05, hep-th/9702008}
\lref\sagnotti{M.Bianchi, hep-th/9702093,ROM2F-97-4}
\lref\bianchi{A. Sagnotti, hep-th/9702098, ROM2F-97-6} 
\lref\sorokin{I.Bandos, K.Lechner, Nurmagambetov, P.Pasti, 
D.Sorokin, 
M.Tonin, hep-th/9701449 }
\lref\tseytlin{A.Tseytlin, Class.Quant.Grav.13:L81-L85 (1996)}
\lref\cvetic{M.Cvetic, A.Tseytlin, Phys.Rev. D53: 5619-5633, 1996}
\lref\polchino{J.Polchinski,NSF-ITP-95-122,hep-th/9510017}
\lref\townsend{P.Townsend,hep-th/9507048}
\lref\azc{J.A.de Azcarraga, J.P.Gauntlett, J.M.Izquierdo, P.K.Townsend,
 Phys.Rev.{\bf D63} (1989) 2443}
\lref\klebanov{S.Gubser,A.Hashimoto,I.R.Klebanov,J.Maldacena
Nucl.Phys.{\bf B472}(1996) 231}
\lref\green{M.B.Green, Phys.Lett.{\bf B223}(1989)157}
\lref\me{D.Polyakov,Nucl.Phys.{\bf B468}(1996)155}
\lref\olive{D.Olive, E.Witten, Phys.Lett.78B:97, 1978}
\lref\bern{J.N.Bernstein, D.A.Leites, Funkts. Analis i ego Pril.,
11 (1977)70-77}
\lref\cederwall{M.Cederwall, A.von Gussich,B.E.W.Nilsson,P.Sundell,
A.Westerberg, hep-th/9611159}
\lref\schwarz{M.Aganagic,J.Park, C.Popescu and J.H.Schwarz,
hep-th/9701166}
\lref\bel{A.Belopolsky, hep-th/9609220}
\lref\witten{E.Witten, hep-th/9610234}
\lref\witte{E.Witten, Nucl.Phys.{\bf B463} (1996), p.383}
\lref\myself{D.Polyakov, LANDAU-96-TMP-4, hep-th/9609092,
to appear in Nucl.Phys.{\bf B484}}
\lref\nberkovits{N.Berkovits, hep-th/9704109}
\lref\berkovits{N.Berkovits, private communications}
\lref\duff{M.J.Duff, K.S.Stelle, Phys.Lett.{\bf {B253}} (1991)113}
\lref\bars{I.Bars, hep-th/9607112}
\lref\grin{M.Green, Phys.Lett. {\bf B223}(1989)} 
\lref\banks{T.Banks, W.Fischler, S.Shenker, L.Susskind, hep-th/9610043}
\lref\seiberg{T.Banks, N.Seiberg, S.Shenker, RU-96-117, hep-th/9612157}
\centerline{\bf 1.Introduction}
  The recent progress in the attempts to explore non-perturbative aspects
of theories of extended objects, such as strings and p-branes, involves
many remarkable features. One of them is the M-theory, the
theory in eleven dimensions whose low-energy limit is the $D=11$
supergravity. It was shown that various formulations of string theories
(such as type I, IIA,  heterotic $E_8{\times}E_8$ or $SO(32)$ and,
not without difficulties, the type IIB ~\refs{\witte})
and various duality symmetries that occur between them 
and between other p-branes  
 follow from M-theory.
Thus the M-theory appears to be a possible candidate  for the underlying
unification theory.
  
 Another development  is the idea of the ``p-brane democracy''
according to which all the extended p-dimensional objects
propagating in a space-time appear on equal ground,playing
equally important role from
the point of view of  non-perturbative physics and strings
are no longer special, with the exception
of the perturbative expansion (see, for instance,~\refs{\townsend}).
Apart from qualitative arguments,  
the practical realization of the idea of p-brane democracy
 is problematic, since no 
systematic quantization is known for p-branes; thus, for instance,
the S-duality for branes is formulated on the level of their
low-energy effective field theories
 (whose actions  may also be rather conjectured  than
rigorously derived). 
  One may consider, however, the different approach which
still regards the theory of open strings as ``more equal than
others'' ~\refs{\sagnotti, \bianchi, \nberkovits}
 even from the non-perturbative
point of view, i.e. the p-brane democracy is complemented
by the ``open-string aristocracy'' (in the terminology 
of ~\refs{\bianchi, \sagnotti}).
The basic idea is that the 
non-perturbative phenomena  in the physics of branes, 
such as dualities may be understood from essentually
perturbative computations
in open string theories. Roughly speaking, it is the presence of the 
worldsheet boundaries that gives rise to the non-perturbative effects.
D-branes are perhaps the best known example of that;
there the massless excitations of an open string with
the ends attached to a hyperplane become the collective coordinates
for the transverse fluctuations of the brane.
 It was shown ~\refs{\polchino} that these objects
are BPS saturated and carry the full set of the Ramond-Ramond charges.
In this paper  we shall consider the particular example of
how the non-perturbative symmetries (dualities) between branes 
 can be understood in terms of the perturbative open-string physics.
Namely,we shall try to show that the well-known S-duality
between strings and fivebranes (which  also follows from 
the M-theory) is closely related to the degenerate case of a
 specific 
gauge symmetry in superstring theory - the picture-changing 
 at zero momentum.

As for the M-theory, unfortunately very few quantitive facts are 
 still known
about its dynamics apart from the low-energy limit.
One particularly promising approach is to try to
formulate  M-theory as a matrix model ~\refs{\banks, \seiberg}.
In this M(atrix) approach the crucial role is played by the
eleven-dimensional SUSY algebra, extended by $p$-form central charges
corresponding to $p$-brane states ~\refs{\azc, \sez, \seiberg}.
More precisely, p-brane solutions give rise to  topological p-form
charges, known as  Page charges, which become generators in the
new extended superalgebras, along with the supercharge and 
the momentum generator.
For instance, the anticommutator of the supercharges becomes
\eqn\lowen{{\lbrace}Q_\alpha,Q_\beta{\rbrace}=
\Gamma_{\alpha\beta}^{\mu}P_\mu + 
{\sum_p}\Gamma_{\alpha\beta}^{\mu_1...\mu_p}
Z_{\mu_1...\mu_p},}
where each (super) p-form charge $Z_{\mu_1...\mu_p}$ in 
the sum corresponds to
(super) p-brane soliton ~\refs{\azc}.
The origin of these charges is supposedly related to
Wess-Zumino terms in the corresponding super p-brane actions
~\refs{\bars}.

This may be illustrated on the example of 
supermembrane solution of $D=11$ supergravity where the expression
for $Z_{\mu_1\mu_2}$ can be found explicitly ~\refs{\duff}.
Indeed, in the $D=11$ supermembrane action the Wess-Zumino term is 
supersymmetric only up to a total derivative term; as a result
integrating this boundary term over the surface of a membrane 
leads to  the presence of the following
following 2-form charge in the SUSY algebra:
\eqn\lowen{Z^{\mu_1\mu_2}=
\int{d^2}\sigma{\epsilon^{0ij}}{\partial_i}X^{\mu_1}
{\partial_j}X^{\mu_2}} 
This integral is nonzero if   
the membrane configuration
defines a non-trivial 2-cycle in eleven dimensions.

As is well-known, the D=11 supergravity, the low-energy limit of the
M-theory, has two non-trivial solutions known as M-branes:
one is the electric-type (the membrane) and another is the magnetic type
(five-brane).These solutions are shown to be related by S-duality.
The $M$-algebra therefore  includes the following
generators: 
${\lbrace}{Q^\alpha},{ P^\mu},{Z^{A_1A_2}},{Z^{A_1...A_5}}\rbrace$,
where $A=(\mu,\alpha)$.There is also a super one-form $Z^A$, which
presence in the M-algebra is quite puzzling. 
Its fermionic part  must be related
to the $\kappa$-symmetry; in the context of Green-Schwarz superstring
theory it has been discussed in ~\refs{\myself}. 
In the $D=10$ context, the presence of the bosonic
component is probably related to the
subtleties of picture-changing at zero momentum(which is a degenerate 
case of 
BRST gauge transformation).Note that, while 
locally the bosonic operator $Z^\mu$ can be 
absorbed into redefinition of the momentum operator , it is not always 
possible globally; and in string theory the operator $Z^\mu$ 
has been shown to play an interesting role in describing the
string winding states ~\refs{\grin}
The (anti)commutation relations of the M-algebra and
 proof that its generators
satisfy Jacobi identities  have been given in ~\refs{\sez}.
For future reference, some of these (anti)commutators are given by:
\eqn\grav{\eqalign{{\lbrace}Q_\alpha,Q_\beta{\rbrace}=\Gamma^\mu_{\alpha
\beta}(P^\mu+Z^\mu) + \Gamma^{\mu_1\mu_2}_{\alpha\beta}Z_{\mu_1\mu_2}+
\Gamma^{\mu_1...\mu_5}_{\alpha\beta}Z_{\mu_1...\mu_5}\cr
\lbrack{P^\mu},Q_\alpha{\rbrack}=\Gamma^\mu_{\alpha\beta}Z^\beta-
{\Gamma^{\mu\nu}_{\alpha\beta}}{Z_{\nu}^\beta}-
{\Gamma^{\mu\mu_1...\mu_4}_{\alpha\beta}}Z_{\mu_1...\mu_4}^\beta\cr
\lbrack{P^{\mu_1}},P^{\mu_2}\rbrack={\Gamma^{\mu_1\mu_2}_{\alpha\beta}}
Z^{\alpha\beta}+{\Gamma^{\mu_1...\mu_5}_{\alpha\beta}}
{Z_{\mu_3..\mu_5}^{\alpha\beta}}\cr
\lbrack{Q_\alpha}Z^\mu{\rbrack}={\Gamma^{\mu}_{\alpha\beta}}Z^\beta,\cr
{...etc.}}}
The covariant $\kappa$-symmetric action for super five-brane has been
recently constructed ~\refs{\witten, \schwarz, \sorokin, \sezgi, \cederwall}
, yet its relation to the 5-form of (3)
is yet to be pointed out.
 In general, the explicit expressions for p-form charges are not
known, as even the classical actions are not yet known for some
super p-branes,let alone more complicated cases of intersecting 
branes.At the same time, it is expected that the M-algebra shall
play an important role in the M-theory and in describing the non-perturbative
physics of extended objects ~\refs{\sez}.
One particular question of interest is what are T- and
S-dualities in the context of extended Poincare superalgebras.
It shall also be of importance to understand the relation between 
 the M-algebra and  D-branes, especially given the role
that D0 - branes are conjectured to play in describing the degrees
of freedom in the M-theory apart from the low-energy limit
~\refs{\banks}.
Recently it has been shown that the SUSY algebra of the (M)atrix
model ~\refs{\seiberg}
does not contain the transverse fivebrane charge,which creates
problems with Lorentz invariance.It is therefore of importance to 
understand the role that the fivebrane plays in the (M)atrix theory.
 In this paper we shall discuss these questions by considering
 chiral reductions of the M-algebra to $D=10$.
The analysis will show a surprizing relation between picture-changing
formalism in string theory and S-duality;  namely we will see
that the picture changing transformation of the D=10 IIB 
Poincare superalgebra gives  rise to  to the 
fivebrane state of the M-theory  and will compute the S-matrix
elements involving this solitonic state.
S-duality shall be interpreted then as a non-perturbative 
counterpart of the
gauge symmetry defined by picture-changing transformations.

\centerline{\bf 2.Duality and Picture - Changing} 
Consider the chiral reduction of the $M$-algebra (3) to $D=10$
which yields the $M$-extension
of the usual (1,0) Poincare superalgebra ~\refs{\sezg}:
\eqn\grav{\eqalign{\lbrace{Q_\alpha},Q_\beta{\rbrace}=
{\Gamma_{\alpha\beta}^\mu}({P_\mu}+Z_\mu)+
{\Gamma^{\mu_1...\mu_5}_{\alpha\beta}}Z_{\mu_1...\mu_5}\cr
\lbrack{P^\mu},Q_\alpha\rbrack={\Gamma^\mu_{\alpha\beta}}Z^\beta -
{\Gamma^{\mu\mu_1...\mu_4}_{\alpha\beta}}{Z_{\mu_1...\mu_4}^{\beta}}\cr
\lbrack{P^{\mu_1}},{P^{\mu_2}}\rbrack={\Gamma^{\mu_1...\mu_5}_{\alpha\beta}}
{Z_{\mu_3...\mu_5}^\beta}\cr
\lbrack{Q_\alpha},{Z^\mu}{\rbrack}=-\Gamma^\mu_{\alpha\beta}{Z^\beta}\cr
\lbrack{Q_\alpha},{Z^{\mu_1...\mu_5}}\rbrack
=-{\Gamma_{\alpha\beta}^{\mu_1...\mu_5}}
Z^\beta + {\Gamma^{\mu_5}_{\alpha\beta}}Z^{\mu_1...\mu_4\beta}\cr
...etc.
}}
Compared with the M-algebra (3), the two-form is now gone and only
one- and five-form charges are left, in agreement with the
S-duality between strings and fivebranes
in the $D=10$ supergravity which is the 
low-energy limit of Green-Schwarz superstring theory.
We are going to show that: 

1)the appearance of the five-form in the
superalgebra (14) follows from picture-changing transformation
of supercharges;

2) the picture-changing transformation
 is to be considered
as a ``generator'' of S-duality.
Consider the charges $Q_\alpha$ of (4) which are given in the 
NSR superstring theory by ~\refs{\shenker}:
\eqn\lowen{Q_\alpha = \oint{{dz}\over{2i\pi}}e^{-{\phi\over{2}}}
\Sigma_\alpha(z)}
Here 
$\phi(z)$ is a bosonized superconformal ghost, $\Sigma_\alpha$
is spin operator for matter fields.
Recall that the  O.P.E. between two $\Sigma$'s is given by
\eqn\lowen{:\Sigma_\alpha:(z):\Sigma_\beta:(w)\sim
{{\epsilon_{\alpha\beta}}\over{{(z-w)}^{5\over4}}}+
{\sum_p}{{\Gamma^{\mu_1...\mu_p}_{\alpha\beta}}\over{{(z-w)}^{{5\over4}-
{p\over2}}}}:\psi_{\mu_1}...\psi_{\mu_p}:(z) + derivatives...}
 where $\psi$'s are NSR worldsheet bosons and the superconformal ghosts
$\beta,\gamma$ are bosonized as
$\gamma=e^{\phi-\chi};\beta=e^{\chi-\phi}\partial\chi$; and
$-<\phi(z)\phi(w)>=<\chi(z)\chi(w)>=log(z-w)$.
Evaluating the anticommutator of two $Q$'s of (5) then gives
\eqn\lowen{\lbrace{Q_\alpha},Q_\beta{\rbrace}=\oint{{dw}\over{2i\pi}}
{1\over{z-w}}\oint{{dz}\over{2i\pi}}{\Gamma^\mu_{\alpha\beta}}
e^{-\phi}{\psi_\mu}(z)=\Gamma^\mu_{\alpha\beta}P_\mu}
where $P_\mu=\oint{{dz}\over2i\pi}e^{-\phi}\psi_\mu$
is an integral of a momentum generator in the $-1$-picture;
i.e. we get the standard result for 
 superalgebra in Green-Schwarz superstring theory
in a flat space-time.
The supercharge of (5)-(7) has been taken in 
the standard ${-{1\over2}}$-picture.
Let us further consider the commutator
$\lbrack{P^\mu,Q_\alpha}\rbrack=\Gamma^\mu_{\alpha\beta}T_\beta$
Here $T_\beta$ is a new fermionic generator ~\refs{\grin, \sez}
We related it to the NSR formulation of the $\kappa$-symmetry generator
in ~\refs{\myself}.The computation using (5) gives
\eqn\lowen{\lbrack{P^\mu, Q_\alpha}\rbrack = 
\Gamma^\mu_{\alpha\beta}\oint{{dz}\over{2i\pi}}
e^{-{3\over2}\phi}\Sigma_\beta\equiv\Gamma^\mu_{\alpha\beta}T_\beta}
It is the integrand of $T_\beta$ that can be shown to generate the
$\kappa$-symmetry transformations in Green-Schwarz superstring theory,
up to picture-changing transformations.
Next, consider the anticommutator $\lbrace{T_\alpha, T_\beta}\rbrace$.
It may be regarded  as the anticommutator of two
$\kappa$-transformations but, more importantly, 
it also has a subtle connection 
directly to the anticommutator of two supercharges in the supersymmetry
algebra, which we shall point out later.The computation gives:
\eqn\grav{\eqalign{\lbrace{T_\alpha, T_\beta}\rbrace=
\oint{{dz}\over{2i\pi}}({1\over2}\Gamma^\mu_{\alpha\beta}
{e^{-3\phi}}\psi_\mu({9\over8}\partial\phi\partial\phi-{3\over2}
\partial^2{\phi})-{3\over2}\partial\psi_\mu\partial\phi+{1/over2}
\partial^2{\psi_\mu}\cr+
\Gamma^{\mu_1\mu_2}_{\alpha\beta}\partial({e^{-3\phi}\psi_{\mu_1}...
\psi_{\mu_3}})+{\Gamma^{\mu_1...\mu_5}_{\alpha\beta}}e^{-3\phi}
\psi_{\mu_1}...\psi_{\mu_5})}}
Let us analyze the r.h.s. of this anticommutator.
Consider the picture-changing gauge transformation, 
defined by the picture-changing operator:
\eqn\lowen{\Gamma_1=:\delta(\gamma)(S_{matter}+S_{ghost}):=
-{1\over2}e^\phi{\psi_\mu}\partial{X^\mu}+e^{2\phi-\chi}b\partial\phi
+e^{\chi}\partial\chi{c}}
Here $S_{matter},S_{ghost}$ are matter and ghost worldsheet supercurrents
(with the superconformal ghosts $\beta$,$\gamma$ being bosonized as
$\gamma=e^{\phi-\chi},\beta=e^{\chi-\phi}\partial\chi$,
$<\chi(z)\chi(w)>=-<\phi(z)\phi(w)>=log(z-w)$).
Then, applying it twice to the sum of the first three terms in the
r.h.s. of the anticommutator, we get the result equal to 
${1\over{16}}\Gamma^\mu_{\alpha\beta}e^{-\phi}{\psi_\mu}$,i.e.
the expression proportional
to the momentum operator in the $-1$-picture.
In other words, the r.h.s of the anticommutator of two fermionic generators
$T$ reproduces the anticommutator $\lbrace{Q,Q}\rbrace$ in the 
non-perturbative version (4)
of the superalgebra  in $D=10$, 
with the five-form (fivebrane) central
term proportional to $e^{-3\phi}\psi_{\mu_1}...\psi_{\mu_5}$,
with the exception of the total derivative three-form term, proportional
to $\sim\partial({e^{-3\phi}}\psi_{\mu_1}...\psi_{\mu_3})$ 
The presence of the total derivative three-form  term 
may be interpreted as the Guven threebrane which, in turn, is nothing but
the intersection of two fivebranes.Thus, one may suspect the correspondence
between total derivative central terms in the super(current) algebra and
intersecting branes.  
 From now on, we will
 concentrate on a fivebrane central term, which appears
to be more fundamental.
In order to understand its relevance to the superalgebra, it
is crucial to point out the connection between the fermionic
generators $T_\alpha$ and the supercharges $Q_\alpha$ of (5),
which determine the regular superalgebra without central terms.
Contrary to what one might suspect, $T_\alpha$ and $Q_\alpha$
are $not$  related  by the picture-changing transformation,
therefore we cannot immediately 
interpret $T_\alpha$ just as a supercharge in
another picture.
Indeed, the naive application of the picture-changing operator
$\Gamma_1$ to $T_\alpha$ gives zero (this is true for the terms
proportional to $S_{matter}$ and to the ghost field $b$, and the
term proportional to $c$ does not apply since $T_\alpha$ is a
contour integral  ~\refs{\shenker, \berkovits}).
 Therefore the relation between $Q_\alpha$ and $T_\alpha$ is rather
subtle.Namely, consider the vertex operator
\eqn\lowen{v_\alpha(k,\bar{k})
R_\alpha = v_\alpha(k,\bar{k})e^{-{3\over2}}\Sigma_\alpha
{e^{ikX}}+ghosts\equiv(\Gamma\bar{k})_{\alpha\beta}{u_\alpha}(k)
e^{-{3\over2}\phi}\Sigma_\alpha{e^{ikX}+ghosts}}
where the ghost terms (not significant for correlation functions
and skipped here) are introduced to insure the BRST invariance.
Here k is the momentum, and the auxiliary momentum $\bar{k}$
is defined so that $(k,k)=0,(\bar{k},\bar{k})=0,(k,\bar{k})=1)$,
i.e. its definition is similar to the one
 for the dilaton vertex operator.
$u_\alpha(k)$ is some constant on-shell space-time spinor.
 The crucial difference, however, is that the polarization tensor
$v_\alpha(k,\bar{k})$ is no longer transverse, i.e. it does not satisfy
the on-shell Dirac equation.Therefore the vertex $R_\alpha$ is
not BRST-invariant for arbitrary $k,\bar{k}$.It $is$ BRST-invariant,
however, in the limit $k\rightarrow{0}$, $\bar{k}\rightarrow\infty$,
and in that limit the picture-changing transformation can be applied.
We will refer to it as a 
``picture-changing in the infinite-momentum frame'', because of the
 mentioned $\bar{k}\rightarrow\infty$ limit.
Therefore, formally we have 
\eqn\lowen{:\Gamma_1\int{{dz}\over{2i\pi}}v_\alpha(k,\bar{k}){R_\alpha}:
=u_\alpha(k)Q_\alpha}
Now, the relation between $T$ and $Q$ is given by

\eqn\lowen{Q_\alpha = :N_{\alpha\beta}T_\beta:}
where the operator $N_{\alpha\beta}$ 
is defined as the ``infinite-momentum''
generalisation of the picture-changing operator $\Gamma_1$ - that is,
for a local fermionic vertex at zero momentum
$V_\alpha(z,k=0)\equiv{V_\alpha(z)}$ such that
$:\Gamma_1V_{\alpha}:=0$,
\eqn\lowen{:N_{\alpha\beta}V_\beta:(z)={{lim}_{\bar{k}\rightarrow{\infty},
k\rightarrow{0}}}(\Gamma\bar{k})_{\alpha\beta}:\Gamma_1{V_\beta}e^{ikX}:(z)}

We see that the operator $N_{\alpha\beta}$, which  essentially is the
special singular limit of the picture-changing transformation,
maps the perturbative SUSY algebra without central charges to the
non-perturbative one,containing the fivebrane.
In the limit $k\rightarrow{0}$ $N_{\alpha\beta}$ effectively
replaces the picture-changing , which is not well-defined at
zero momentum ~\refs{\shenker}.
 Before going further, let us comment on
 what appears to be a paradox - the fact
that the anticommutator of two kappa-transformations 
gives something like the supersymmetry algebra.
It is well-known that $\kappa$-symmetry algebra is 
quite different from the superalgebra,i.e. the combination of
two kappa-transformations is again a kappa-transformation.
At first glance, it is not consistent with the relation (9)
The answer,however,
 is that that the structure constants in the $\kappa$-symmetry
algebra depend on superspace coordinates,namely 
they are proportional to $\partial\theta_\alpha$  where
$\theta_\alpha$ is a GS variable,corresponding to
$\sim{e^{\phi\over2}}\Sigma_\alpha$ in the NSR formalism.
The O.P.E of this
 fermionic structure constant with the $\kappa$-symmetry generator
must be then normally ordered,
giving rise to the 
$bosonic$ generator,proportional to the picture-changed momentum plus
central terms. 
Let us now return to the five-form central term in the anticommutator (9),
which defines the ``fivebrane emission vertex'' at zero momentum.
This open-string vertex is rather peculiar since it does not seem to
describe the emission of any massless particle in the theory of
open strings, at the same time it is not BRST trivial.
Also, the ghost terms must be added to it in order to make its
BRST-invariance manifest; however, because these terms are cumbersome and 
because they do not affect correlators, we will skip them.
There is no analogue of this vertex in a picture with ghost number
zero, which is quite an unusual situation for bosonic vertex operators.
In other words, the essentially non-zero ghost number is a crucial factor
for such a ``brane emission vertex''.
It has, however, the picture-changed counterpart in the $+1-picture$.
Recall that the higher picture-changing operators of ghost number $n$
$\Gamma_n$
(in the perturbative string theory,
$\Gamma_n=:{(\Gamma_1)^n}:+\lbrace{Q_{BRST},...}\rbrace$ for
physical vertex operators at non-zero momentum)
are given by
\eqn\grav{\Gamma_n=e^{n\phi}S\partial{S}...\partial^{n-1}S}
where $S=S_{matter}+S_{ghost}$ is the sum of matter and ghost
worldsheet supercurrents.Then 
\eqn\lowen{:\Gamma_4{e^{-3\phi}}\psi_{\mu_1}...\psi_{\mu_5}:=
e^\phi{\psi_{\mu_1}}...{\psi_{\mu_5}}}
Of course,
this alternative formulation of the five-form term 
has no analogue in the 0-picture as well.Analogously, the 
counterpart of the three-brane total derivative central term
in the $+1$-picture is given by 
$\partial({e^\phi}\psi_{\mu_1}...\psi{\mu_3})$. One may check that
\eqn\lowen{\lbrace:{\Gamma_2}T_\alpha:,:{\Gamma_2}T_\beta\rbrace=
\Gamma^\mu_{\alpha\beta}:\Gamma_2{e^{-\phi}}\psi_mu:+
\Gamma^{\mu_1...\mu_3}_{\alpha\beta}
\partial(e^{\phi}\psi_{\mu_1}...\psi_{\mu_3})
+\Gamma^{\mu_1...\mu_5}_{\alpha\beta}e^{\phi}\psi_{\mu_1}...\psi_{\mu_5}}
For the sake of convenience,
from now on we will be using the version of the five-form term
in the $+1$-picture in our analysis, since its properties
are quite equivalent to its $-3$-picture counterpart.In general,
any brane emission vertex in a negative picture appears to have
the equivalent counterpart in the non-negative picture,even though
 picture-changing for such non-perturbative vertices involves
many subtleties.It is therefore enough to consider non-negative pictures
to study the properties of such brane vertex operators.
As we have already mentioned,
the remarkable property of the five-form operator is that it cannot
be connected with any operator of  ghost number zero by means of
picture-changing.
Namely, the following formal consideration may be given.
Let $\lbrace{V_n}\rbrace$ be the set of physical states (or the set
of vertex operators) having ghost number $n>0$.Let us further
define $\lbrace{\tilde{V_n}}\rbrace$ as the set of those
vertices of ghost number $n$ which can be obtained as a result of 
picture-changing transformation of vertices of lower 
non-negative ghost numbers,i.e.

$\lbrace{\tilde{V_n}}\rbrace\supset\Gamma_1\lbrace{n-1}\rbrace\oplus
\Gamma_2\lbrace{V_{n-2}}\rbrace\cap\Gamma_1\lbrace{V_{n-1}}\rbrace...\oplus
\Gamma_n\lbrace{V_{0}}\rbrace\cap\Gamma_{n-1}\lbrace{V_{1}}\rbrace...\cap
\Gamma_1\lbrace{V_{n-1}}\rbrace$ 
where $:\Gamma_n:$ are picture-changing operators of higher
ghost numbers.
 
Consider the cohomology classes :
\eqn\lowen{H_n={{\lbrace{V_n}\rbrace}\over{\lbrace{\tilde{V_n}}\rbrace}}}
The vertex $V(z,k)$ of (16) is then the element of $H_1$.
In general, we expect all the perturbative string states
to belong to the ``elementary'' ghost number cohomology $H_0$,
while cohomologies of non-zero ghost numbers account for the
non-perturbative physics ,i.e. branes. 
Note that, while it is of a conformal dimension 1,
the charge $Z_{\mu_1...\mu_5}$
 should , in principle,
be an integral over the volume of a fivebrane of some expression
of a dimension 5, in connection with  Wess-Zumino term for a fivebrane.
 We conjecture therefore that the five-form 
$e^{-3\phi}\psi_{\mu_1}...\psi_{\mu_5}$
in the  algebra (9),which is
expressed in terms of open string
variables , is essentially of a D-brane origin.The state
$V(z,k=0)|0>$ should be related to the boundary state of a 
D5-brane, where $|0>$ is an open string vacuum.
 Indeed, computing various S-matrix elements of the vertex operator
$V(z,k)=e^\phi{\psi_{\mu_1}}...{\psi_{\mu_5}}e^{ikX}$ in open
string theory one may check that they all vanish, which
is not surprizing 
since this operator does not describe the emission of any massless
particle in the perturbative string spectrum.
In the presence of D-branes , however, the S-matrix elements
of $V(z,k)$ are nonzero, due to its interaction with Ramond-Ramond charges.
 In fact, 
$Z_{\mu_1...\mu_5}$  defines a solitonic-type state which interacts
with Ramond-Ramond (RR) charges of a D-brane. 
The computation gives the nonzero S-matrix elements
$<V(z_1) V^{RR}(z_2, k_1) V^{RR}(z_3, k_2)>$ and
$<V(z_1, k_1)V^{RR}(z_2, k_2)V^{NS-NS}(z_3, k_3)>$ on a disk
with mixed Dirichlet - Neumann boundary conditions which
determine the scattering off D-brane with the inserted
boundary state (16)(in the first of them the vertex $V(z_1)$
must be multiplied by an anti-holomorphic part).
The first one is just the analogue of the p-brane 
gravitational lensing of ~\refs{\klebanov}; while the 
second describes an interesting physical process -
the Ramond-Ramond particle becoming the NS-NS particle
due to interaction with the monopole-like state (16).
In this paper we shall present the computation of
 the matrix element corresponding to this process. 
The vertex operators in the appropriate pictures must have a total
left $+$ right ghost number $-2$ in order to to compensate the ghost number
anomaly:
\eqn\grav{\eqalign{
V^{RR}(z,\bar{z},k)= e^{-{1\over2}(\phi+\bar\phi)}\Sigma_\alpha
{{\bar\Sigma}_\beta}\Gamma^{\nu_1...\nu_p}F_{\nu_1....\nu_p}(k)e^{ikX}\cr
V^{NS-NS}(z,\bar{z},k)=
e^{-\phi-\bar\phi}\psi^\mu{\bar\psi^\nu}\rho_{\mu\nu}(k)
e^{ikX},\cr
V(z,k)= e^\phi{\psi_{\mu_1}}...{\psi_{\mu_2}}
{{{\tilde{Z}}^{\mu_1...\mu_5}}}.}}
The ``tilde'' in the polarization tensor $\tilde{Z}$
 of $V(z,k)$ is used to avoid
a confusion with the $Z$ of (16).
Note that, as it has been explained in ~\refs{\klebanov},
we may consider the correlation function on the half-plane
first and then, if needed,  map it conformally to the disk.
Then, extending the (anti)holomorphic fields to the whole
complex plane, we get the following relations between holomorphic
and antiholomorphic parts reproducing all the O.P.E.'s on the half-plane:
\eqn\grav{\eqalign{\bar\psi^\mu(\bar{z}) = \pm{\psi^\mu}(\bar{z})\cr
\bar{X^\mu}(\bar{z}) = \pm{X^\mu}(\bar{z})}}
The sign depends on whether there are Neumann or Dirichlet
boundary conditions for  particular $\mu$'s;
and
\eqn\lowen{\bar\Sigma_\alpha(\bar{z})=M_{\alpha\beta}\Sigma^\beta(\bar{z})}
where $M_{\alpha\beta} = (\Gamma^0...\Gamma^p)_{\alpha\beta}$.
 We now have to consider:
\eqn\grav{\eqalign{<V^{RR}(z,\bar{z},k_1)V^{NS-NS}(w,\bar{w},k_2)
V(0,k=0)>=
{(F^{(p)}M)^{\alpha\beta}}(k_1)B^{\mu\nu}(k_2)
{\tilde{Z}}^{\mu_1...\mu_5}\times\cr
\times <e^{-{\phi\over2}}{\Sigma_\alpha}(z)e^{-{{\bar\phi}\over2}}
{\bar\Sigma_\beta}(\bar{z})e^{-\phi}\psi_\mu(w)e^{-\bar\phi}
\bar\psi_\nu(\bar{w})e^{\phi}\psi_{\mu_1}...\psi_{\mu_5}(0)>
\times\cr\times
<e^{i({k_1}X)}(z_1,{\bar{z}}_1)e^{i({k_2}X)}(z_2,{\bar{z}}_2)>}}
Here $F^{(p)}(k)= \Gamma^{\nu_1...\nu_p}F_{\mu_1...\mu_p}(k)$ is
the Ramond-Ramond field strength contracted with the antisymmetrized
product of gamma-matrices , $B^{\mu\nu}(k)$ is the axion's polarization
tensor and ${\tilde{Z}}^{\mu_1...\mu_5}$ is  the 
topological charge associated
with the boundary vertex $e^\phi{\psi_{\mu_1}}...{\psi_{\mu_5}}$.

While  the correlator of two exponents of $X^\mu$'s in (22)
(the four-point function) is easy to compute, some work is
needed to calculate the five-point function involving the
 fermions, ghosts and spin fields.
This five-point correlator is computed by
 applying the O.P.E. rules involving spin fields and worldsheet
fermions and using the theorem about the poles of meromorphic function.
The result of the computation is given by:
\eqn\grav{\eqalign{<e^{-{\phi\over2}}\Sigma_\alpha(z)
e^{-{\bar\phi\over2}}\Sigma_\beta(\bar{z})e^{-\phi}\psi^\mu(w)
e^{-\bar\phi}\bar\psi^\nu(\bar{w}){e^\phi}\psi^{\mu_1}...\psi^{\mu_5}(0)>
={{{2(\Gamma^\mu\Gamma^\nu\Gamma^{\mu_1...\mu_5})}_{\alpha\beta}}\over
{(z-w)(w-\bar{w})^2{{\bar{z}}^2}}}\cr+
{{{(\Gamma^\mu\Gamma^{\mu_1...\mu_5}
\Gamma^\nu)}_{\alpha\beta}}\over{(z-w)w(\bar{z}-\bar{w})\bar{z}\bar{w}}}
+{{{(\Gamma^\mu\Gamma^\rho\Gamma^{\mu_1...\mu_3})}_{\alpha\beta}
g^{\mu_4\nu}g^{\mu_5\rho}+permut.(\mu,\nu;\mu_1...\mu_5)}\over
{(z-w)(w-\bar{w})(\bar{z}-\bar{w})\bar{z}\bar{w}}}+\cr
{{{(\Gamma^\mu\Gamma^{\mu_1...\mu_5}\Gamma^\nu)}_{\alpha\beta}}\over
{(z-w)(\bar{z}-\bar{w})(w-\bar{w})\bar{z}\bar{w}}}+
{{{(\Gamma^\mu\Gamma^\rho\Gamma^{\mu_1...\mu_3})}_{\alpha\beta}
+permut.(\mu,\nu;\mu_1...\mu_5)}\over{(z-w)\bar{z}(w-\bar{w})w\bar{w}}}
+\cr+
{{{2(\Gamma^\nu\Gamma^\mu\Gamma^{\mu_1...\mu_5})}_{\alpha\beta}}\over
{(z-\bar{w})(w-\bar{w})^2{\bar{z}}^2}}+
{{{(\Gamma^\nu\Gamma^{\mu_1...\mu_5}\Gamma^\mu)}_{\alpha\beta}}\over
{(z-\bar{w})\bar{w}(\bar{z}-w)\bar{z}w}}+
{{{(\Gamma^\nu\Gamma^{\mu_1...\mu_5}\Gamma^\mu)}_{\alpha\beta}}\over
{(z-\bar{w})(\bar{z}-w)(\bar{w}-w)w\bar{w}}}+\cr+
{{{(\Gamma^\nu\Gamma^{\mu_4}\Gamma^{\mu_1...\mu_3})}_{\alpha\beta}
g^{\mu_5\mu}+permut.(\mu,\nu;\mu_1...\mu_5)}\over{(z-\bar{w})(\bar{w}-w)
\bar{z}w}}({1\over{\bar{z}-w}}+{1\over{\bar{w}}})+\cr+
{{{(\Gamma^{\mu_1...\mu_5}\Gamma^\mu\Gamma^\nu)}_{\alpha\beta}}\over
{{{z}}^2(w-\bar{w})}}({1\over{{{\bar{z}}-\bar{w}}^2}}+
{1\over{(z-\bar{w})(w-\bar{w})}})+
{{{(\Gamma^{\mu_1...\mu_5}\Gamma^\nu\Gamma^\mu)}_{\alpha\beta}}\over
{z^2(w-\bar{w})(\bar{z}-\bar{w})^2}}+\cr
({{{(\Gamma^{\mu_1...\mu_3})}_{\alpha\beta}g^{\mu\mu_4}g^{\nu\mu_5}}\over
{z\bar{z}(\bar{z}-w)(\bar{z}-\bar{w})(w-\bar{w})}}+
{{{(\Gamma^{\mu_1...\mu_3}\Gamma^{\mu_4}\Gamma^{\mu_5})}_{\alpha\beta}
g^{\mu\nu}}\over{z{(w-\bar{w})^2}{\bar{z}^2}}}+
{{{(\Gamma^{\mu_1...\mu_3}\Gamma^{\mu_4}\Gamma^{\mu_5}\Gamma^{\mu}
\Gamma^{\nu})}_{\alpha\beta}}\over{z(\bar{z}-\bar{w})w\bar{w}(\bar{w}-w)}}
+\cr+
{{{(\Gamma^{\mu_1...\mu_3}\Gamma^{\mu_4}\Gamma^{\mu_5}\Gamma^{\nu}
\Gamma^{\mu})}_{\alpha\beta}}\over{z(\bar{z}-w)w\bar{w}(w-\bar{w})}}
+{{{(\Gamma^{\mu_1...\mu_3}\Gamma^{\mu_4}\Gamma^{\mu_5}\Gamma^{\mu\nu})}_
{\alpha\beta}}\over{z(w-\bar{w})(\bar{z}-\bar{w})\bar{z}\bar{w}}}+
perm(\mu,\nu;\mu_1...\mu_5))+
\cr+
{{{(\Gamma^{\mu\nu}\Gamma^{\mu_1...\mu_5})}_{\alpha\beta}}\over
{(w-\bar{w}){(z-\bar{w})^2}{\bar{z}^2}}}-
{{{(\Gamma^{\mu\nu}\Gamma^{\mu_1...\mu_5})}_{\alpha\beta}}\over
{2(w-\bar{w})(z-\bar{w})(\bar{z}-\bar{w})\bar{z}\bar{w}}}+
(z\longleftrightarrow{\bar{z}},\alpha\longleftrightarrow\beta)}}
After putting  together all the factors in (22)
we obtain the final expression for the matrix element corresponding
to the process of the NS-NS particle producing
the Ramond-Ramond state due to the interaction 
with the Olive-Witten monopole (16):

\eqn\grav{\eqalign{<V^{RR}(z,\bar{z},k_1)V^{NS-NS}(w,\bar{w},k_2)
V(0)>=
{{(z-\bar{z})^{2{{k_1^\parallel}^2}}(w-\bar{w})^{2{{k_2^\parallel}^2}}
(z-w)^{(k_1k_2)}}
\over{{(z-\bar{w})}^{{k_1^\perp}{k_2^\perp}-{k_1^\parallel}{k_2^\parallel}}}}
\times\cr\times
{\lbrack}Tr(F(k_1)MB(k_2)\tilde{Z})G_1(z,\bar{z},w,\bar{w})+
Tr(F(k_1)M\tilde{Z}B(k_2))
G_2(z,\bar{z},w,\bar{w})+\cr+Tr(F(k_1)M\Gamma^\mu{\tilde{Z}}\Gamma^\nu)
B_{\mu\nu}(k_2)G_3(z,\bar{z},w,\bar{w})+\cr+
Tr(F(k_1)M\Gamma^\mu\Gamma^{\mu_1...\mu_4}){{\tilde{Z}}_{\mu_1...\mu_5}}
B_{\mu\mu_5}(k_2)G_4(z,\bar{z},w,\bar{w}){\rbrack}
\delta({k_1^\parallel}+{k_2^\parallel})}}
Here
$k_{(1,2)}^\parallel$ and $k_{(1,2)}^\perp$ are the 
longitudinal and transverse components
of the $k^\mu$'s with respect to the
D p-brane;while the matrices
$F(k_1)=\Gamma^{\mu_1...\mu_p}F_{\mu_1...\mu_p}(k_1)$,
$B(k_2)=\Gamma^{\mu\nu}B_{\mu\nu}(k_2)$ and
$\tilde{Z}=\Gamma^{\mu_1...\mu_5}{\tilde{Z}}_{\mu_1...\mu_5}$
are the polarization tensors contracted with 
 antisymmetrized $\Gamma$-matrices. The functions
$G_i(z,\bar{z},w,\bar{w})$,$i=1,...4$ are given by:
\eqn\grav{\eqalign{G_1(z,\bar{z},w,\bar{w})=
{2\over{(z-w)(w-\bar{w})^2{\bar{z}^2}}}+{2\over{(z-\bar{w})(w-\bar{w})^2
{\bar{z}}^2}}+{1\over{(w-\bar{w})(z-\bar{w})^2{\bar{z}}^2}}-\cr-
{1\over{2(w-\bar{w})(z-\bar{w})(\bar{z}-\bar{w})\bar{z}\bar{w}}}+
(z\longleftrightarrow{\bar{z}});}}
\eqn\grav{\eqalign{G_2(z,\bar{z},w,\bar{w})=
{1\over{z^2(w-{\bar{w}})^2}}({1\over{{\bar{z}}-w}}-{1\over{\bar{z}
-\bar{w}}})+\cr+
{1\over{\bar{z}w\bar{w}(w-\bar{w})}}({1\over{z-w}}-{1\over{z-\bar{w}}})
+(z\longleftrightarrow{\bar{z}});}}
\eqn\grav{\eqalign{G_3(z,\bar{z},w,\bar{w})=
({1\over{(z-\bar{w})(\bar{z}-w)w\bar{w}}}+{1\over{(z-w)(\bar{z}-\bar{w})
w\bar{w}}})
({1\over{\bar{z}}}
+{1\over{w-\bar{w}}})+\cr+
(z\longleftrightarrow{\bar{z}})}}
\eqn\grav{\eqalign{G_4(z,\bar{z},w,\bar{w})=
{1\over{(w-z)(w-\bar{w})\bar{z}\bar{w}}}({1\over{\bar{z}-\bar{w}}}+
{1\over{w}})+\cr+
{1\over{(z-\bar{w})(w-\bar{w})\bar{z}}}({1\over{z(z-w)}}-
{1\over{w(\bar{z}-w)}})+(z\longleftrightarrow{\bar{z}})}}
This concludes the computation of the correlation function (22).
\vskip 0.5in
\centerline{\bf 3.Conclusion}
We have shown that the 5-form topological term,
whose presence in the non-perturbative
Poincare superalgebra for type IIB strings  
must follow from the M-theory, is obtained as a result of 
a special gauge transformation of the $perturbative$
open string superalgebra, which the picture-changing at zero
momentum is. This is consistent with the idea of 
``open string aristocracy'' ~\refs{\bianchi, \sagnotti}
which implies that many aspects 
of the non-perturbative physics  of branes, such as
S-duality in our case, in fact may be hidden in the perturbative structure
of open strings.In the example that we have presented in this paper,
 the string-theoretic analogue of the monopole part in the
Olive-Witten's result for field theory ~\refs{\olive}
 has
appeared as a consequence of the singularity of picture-changing
at $k=0$ (I am grateful to Augusto Sagnotti for pointing
out to me this connection).
Alternatively, one may think of the picture-changing operator 
at zero momentum (or, equivalently, in the infinite momentum frame) 
 as effectively creating  boundaries on the worldsheet which, in turn, 
 give rise to various non-perturbative effects.
This is in accordance with the apparently D-brane nature 
of the 5-form charge in the superalgebra (9).
 It would certainly be interesting to explore the connection between
these two viewpoints, as well as to understand in more details
how the presence of the boundaries modifies current algebras in general.
This may help us to  understand better the still obscure
relation between
 p-brane solutions of low-energy effective theories
 and p-brane dynamics in general, by using the interplay between
the physics in the target space and on the worldsheet (worldvolume).
For instance, due to the relation between 
target space and worldsheet metrics:
\eqn\lowen{\gamma_{ij}={\partial_i}X^\mu{\partial_j}X^\nu{G_{\mu\nu}(X)}}
a singularity in the metric of a given p-brane solution
shall induce a singularity on the worldsheet which,in turn,
one may try to
represent as an insertion of some boundary vertex operator.
P-brane dynamics shall be explored then by studying the properties of the
insertion.We hope to address this question in future papers; 
it would be especially interesting to apply this program
to the case of intersecting branes since their classical action
is  yet to be constructed.
Next, recall that  the monopole-like state (16)
which we have also associated with the boundary state of a D5-brane
appears to be an element of a  ``ghost-number''
cohomology  class $H_1$  defined in (18).
One may expect that BRST non-trivial elements of higher
cohomology classes $H_n$'s should somehow contain the
non-perturbative part of an open-string  spectrum (D-branes)
and dualities shall then appear as mappings between
these cohomology groups.
 Studying the properties of these cohomologies shall involve
 many subtleties connected with the picture-changing
and it requires certain accuracy. 
  
We conclude by giving some intuitive arguments that
emphasize the connections between picture-changing, ghost
number cohomologies and S and T- dualities.
As is well known, the  total ghost number $n$ of a correlation function
(corresponding to some term in the low-energy effective action)
and the genus  $g$ of the worldsheet 
( the expectation value of the dilaton)
are related by the constraint
\eqn\lowen{n=2(1-g).}
At the same time, the picture-changing at zero momentum essentially
generates the map between the ghost number cohomologies $H_n$.
Due to (30), a map between the $H_n$'s should
correspond to the transformation of the dilaton field,
ant this is where the relation with the S-duality shall appear.
One may also anticipate an important role of the Ramond-Ramond vertices
in the relation between S-duality and ghost number cohomologies
due to their property to effectively change the genus of the worldsheet
~\refs{\tseytlin, \cvetic}
  As to the connection with  $T$-duality, consider the
space of pseudodifferential superforms $\Lambda^{r|s}$ with
$r$ and $s$ being even and odd degrees respectively ~\refs{\bern}
(these forms may also be understood as
the extended objects of the dimension $r|s$
propagating in a target space ~\refs{\bel}).  
While the usual differentiation changes the even degree $r$ only,
the odd degree $s$ is shifted by the operation analogous to the
picture-changing , i.e. the dimension of
the extended object is shifted by the transformation of the 
picture-changing: 
\eqn\lowen{\Gamma_1: \Lambda^{r|s}\rightarrow{\Lambda^{r|s-1}}}
On the other hand, $T$-duality, which exchanges
the Dirichlet and Neumann boundary conditions on the ends of
an open string transforms a $D$-brane into either $(D+1)$- or 
$(D-1)$-brane ,depending on whether the $T$-duality is performed
in  transverse or tangent directions with respect to
the $D$-brane. To make this argument precise, we  need 
to understand  the relation between the string-theoretic
picture-changing formalism and the one involving the superforms,
in more accurate terms.

\centerline{\bf Erratum}
The previous version of this paper contained the incorrect
computation of the anticommutator of two supercharges in the
$+{1\over2}-picture$.We claimed to have obtained the 5-form central
term (16) in the anticommutator.However, the accurate calculation
shows that this term appears to be proportional to the matrix factor
$\Gamma_\mu\Gamma^{\mu_1...\mu_5}\Gamma^\mu$, which vanishes in 
$D=10$ - and this factor was missing in the previous version.
Therefore, the central terms  in the superalgebra
do not appear under the naive picture-changing, contrary to
what was previously suggested.
The mistake in the computation led to the erroneous  assumption
that the non-perturbative central terms in the superalgebra
may be obtained just by applying the usual picture-changing transformation
to the perturbative superalgebra (7).However, as it was shown here,
the relation between perturbative superalgebra (7)and the one containing
the fivebrane requires the serious modification of the picture-changing
transformation.This modification , described in (14)
is referred to as ``picture-changing in the infinite momentum frame''.
This name suggests certain relevance to the matrix theory, which we 
hope to discuss in the upcoming paper.
The author is very thankful to N.Berkovits and E.Witten,
whose kind remarks have
pointed out the  computational error 
contained in the previous 
version of this paper,  leading to incorrect physical interpretations. 
  
\centerline{\bf Acknowledgements}
This work is supported by the Grant of the European Post-Doctoral
Institute (EPDI).The author is grateful to the
Institut des Hautes Etudes Scientifiques and especially to
Thibault Damour for the hospitality. 
 The final stage of the present work has been completed 
during the author's visit to the Universita di Roma 2
``Tor Vergata''.
It is a pleasure to thank Massimo Bianchi, Augusto Sagnotti
and the whole High Energy group in Tor Vergata
for the special hospitality they provided.
The author thanks
 C.Angelantoni, M.Bianchi, N.Hambli,
G.Pradisi, A.Sagnotti and Ya.S.Stanev for illuminating discussions.
I'm grateful to A.Sagnotti for pointing out to me the connection 
with the Olive-Witten's result.
I'm thankful to N.Berkovits and E.Witten who have pointed out
to me the error in the previous version. 
\listrefs
  
\end